\documentclass[aip,jap,aps,prb,twocolumn,reprint,superscriptaddress,showpacs,nofootinbib]{revtex4-1}
\pdfoutput=1

\usepackage[utf8]{inputenc}
\usepackage{graphicx}
\usepackage{subfigure}
\usepackage{filecontents}
\usepackage{amsmath}
\usepackage{amssymb}
\usepackage{bm}
\usepackage{color}

\usepackage{latexsym}
\usepackage{dcolumn}
\usepackage{amsxtra}

\def\mb{\mbox}

\def\mb{\mbox}

\begin{document}
\preprint{}

\title{Time-Energy Uncertainty Limit for spin-related wavepacket evolution}

\author{G. Bonfanti}
\affiliation{Grupo de Desarrolo e Inversiones,CDMX, M\'{e}xico}

\author{L. Diago-Cisneros}
\affiliation{Facultad de F\'{\i}sica, Universidad de La Habana, Cuba}

\date{\today}

\begin{abstract}
In this report we study the quantum transport of charge carriers for low dimensional systems with spin-orbit coupling by means of Heisenberg's inequalities. To develop our analysis, an accurate \emph{gendanken} experiment was carefully put together, mainly based on the spin-field effect transistor phenomenology and taking into account several wide accepted approaches on quantum mechanical limited determinism. While verifying the applicability of time-energy uncertainty relation (TEUR) during electronic wavepacket's evolution through a semiconductor quantum wire, some qualitative information related the dynamic behavior of the system is found. The problem is also approached in the framework of the stationary phase method to guarantee robust coherence for wavepacket's evolution, which lately implies certain restrictions on the incident energy values for the envisioned cases. Tailoring different input parameters such as: incoming particles spin polarization, barrier thickness and Rashba's spin-orbit interaction (R-SOI) strength, we have observed appealing effects while the packet evolves through a quasi-1D heterostructure. Hopefully some of them could be of help in spintronics device designing. We have obtained most of the numerical simulation results, within the so-called conservation zone. However, for certain barrier thicknesses, some values drop into the forbidden area regarding the clear theoretical limit imposed by the TEUR. As an expected bonus throughout our theoretical validation of the TEUR, spin splitting due to finite values of R-SOI was nicely noticed.
\end{abstract}


\maketitle

\section{Introduction}
 \label{Prerem}

The time energy uncertainty relationship (TEUR) is one of the aspects around physics that shows and actual and relevant controversy \cite{Giribet,Olkhovsky}. The debate arises from the nonexistence of an operator for time in Quantum Mechanics making the deduction of a TEUR formally impossible. Despite the fact that the existence of the TEUR has been ended by Aharonov and Bohm \cite{Ahoronov} by stating that this relationship must not ever be consider as a general implication of Quantum Mechanics it can be stated that it presents some applicability in some areas of solid state physics. Such areas consider the existence of limitations in the simultaneous measurement between time and energy, better said, some approximations are made in order to be able to deduce and use a TEUR leading to practical results of scientific relevance. Those applications are found in particle physics, nuclear physics, gravitation \cite{Aharonov}, Measurement theory \cite{Aharonov}, the non relativistic tunnel effect \cite{Olkhovsky}, etc.
In the last years the manipulation of the spin has initiated an area of research in the field of nano-technology know as spintronics. This initiation is due to the impressive technological improvements presumed by spintronics devices. Because spintronics devices codify data by using the orientation of the electron's spin and this spin changing operation consumes small amounts of energy it's believed that spintronics gadgets will require little weight batteries, will have the ability of being disconnected between operations becoming non volatile memories, will generate minimum heat by using the little increment in the spin changing operation.
The applicability shown by different authors lies on the construction of an adequate gedanken experiment which will assure the correct interpretation of the data obtained. The main porpoise of this report is to use the TEUR in order to obtain quantitative (semi-quantitative) information on the dynamical behavior of an electronic wavepacket with Rashba's spin orbit interaction (R-SOI), during its temporal evolution through a semiconductor structure, all of this by forming the correct gedanken experiment. We use the stationary phase method (SPM) \cite{Bohm} which has been able to describe quantum transport as other techniques in the area\cite{Leo1,Coppola}. The present study may favor the design of electronic devices by quantitatively showing how different energy zones can be selected by breaking the degeneracy state of the electrons by the spin.

\section{Physical system and \emph{gedanken} experiment}

\begin{figure}
 \hspace*{-8mm}\includegraphics[width=.95\linewidth]{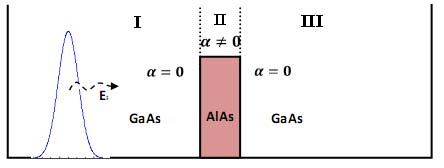}\\
 \caption{\label{Fig0B} Lateral graphic representation of the electronic wave packet evolution trough alternating constant-density regions of a Q1D semiconducting system. In the region II had been induced a R-SOI, whose finite strength $\alpha_{\mb{\tiny R}}$ --$\alpha$ for simplicity--, is tuned \textit{via} a gate voltage.}
\end{figure}

The device under study is a one dimensional Datta and Das spin field effect transistor \cite{Datta}. On the surface above the heterostructure $GaAs/AlAs/GaAs$ two metal plates are placed and a voltage is induced in order to repel the electrons and produce a quasi one dimensional electron gas which can be used to inject a polarized spin current of electrons trough the heterostructure. The physical system can be represented as the band profile shown in Fig.\ref{Fig0B}. In our setup's propagation medium, two $GaAs$-electrodes were taken as emitter and collector of wave packets (Regions I and III, respectively, in Fig.\ref{Fig0B}). The embedded $AlAs$-barrier (Regions II in Fig.\ref{Fig0B}) is capable to induce R-SOI electrostatically. Additionally, we suppose two metal plates placed on top, whose induced voltage repels the electrons from is a quasi-two-dimensional (Q2D) electron gas producing a Q1D-biased quantum wire as a new propagation medium, which now can transport electric and spin currents on an electronic wavepackets trough the Q1D system [see Fig.\ref{Fig0B}]. In order to achieve a successful interpretation of the data a \emph{gedanken} experiment must be constructed by using three different interpretations of the TEUR and collecting from them uppermost components that better support the envisioned problem. All of this done with the purpose of studying the temporal evolution of an electronic wavepacket that interacts with a potential barrier incorporating R-SOI by measuring instantaneously and individually the total energy of the system.

\subsection{Different interpretations of the TEUR}

Mandelstamm and Tamm \cite{Aharonov}
\vspace{3mm}\\
The first one to be used is the Mandelstamm and Tamm interpretation. By using it we obtain that $\Delta E$ indicates the energy range where we can appreciate the statistic weight of the polarized spin states of the electrons. Meanwhile $\Delta t$ indicates the time range where the packet doesn't suffer any significant spreading. From this interpretation we use the observation time with the Stationary Phase Method \cite{Bohm} in order to guarantee a insignificant spreading.
\vspace{3mm}\\
Landau, Peierls and Lifshitz \cite{Aharonov}
\vspace{3mm}\\
This interpretation is followed by the one developed by Landau, Peierls and Lifshitz. They suggest taking into consideration the collision of two particles, the observed one and the pattern one. All done with the goal of eliminating the statistical weight carried in the interpretation give by Mandelstamm and Tamm. The observed particle is the point decided to follow of the wavepacket, the pattern particles is the potential barrier. Therefore the statistical weight is eliminated obtaining an individual and instantaneous measurement of the total energy presented by the wavepacket. However this interpretation uses perturbation theory considering the potential barrier time dependent with second order terms.  Because we are dealing with a stationary potential and we are not using perturbation theory another interpretation must be added to our gedanken experiment.
\vspace{3mm}\\
Fock and Krilov \cite{Aharonov}
\vspace{3mm}\\
The third on is the Fock and Krilov interpretation, where we make the assumption that one of our particles acts as a clock, in this case the centroid of the wavepacket. The velocity with which the centroid is moving is the group velocity of the wavepacket, therefore to ensure the act of our centroid we need to guarantee that the group velocity of the wavepacket is greater than the velocity of our pattern particle (the barrier) before, during, and after the collision. Because a stationary barrier is being used we eliminate the time-dependent potential carried by the perturbation theory used before and an individual treatment of the particles is obtain instead of statistical fluctuations coming from the wave function associated to the system.
\vspace{3mm}\\
Bohr's Postulate
\vspace{3mm}\\
Finally we use Bohr's postulate which mentions that the limited determinism of indirect measurements may be deduced from the mathematical formalism of the statistical uncertainties stating that essentially the statistical uncertainties and the individual uncertainties must be the same.
Using these interpretations we construct a gedanken experiment making possible the study of physical phenomena where the observables have a statistical meaning by using individual measurements.

\section{Hamiltonian for Rashba's spin orbit coupling}

Datta and Das have shown previously \cite{Datta} that the expression for the hamiltonian needed to analyze the behavior of an electronic wavepacket in a one dimensional box of length $L$ is
\begin{equation}
\hat{H}=\hat{H}_{0}+\hat{H}_{so},
\end{equation}
with
\begin{equation}
\hat{H}_{0}=-\frac {\hbar^{2}} {2m^{*}} \frac {d^{2}} {dy^{2}}+V(y),
\end{equation}
and
\begin{equation}
\hat{H}_{so}=H_{so}(x)+H_{so}(y).
\end{equation}
The first term of the latter equation can be neglected obtaining \cite{Burgos}
\begin{equation}
 \label{Rashba}
\hat{H}_{so}=-i\alpha_{R}\sigma_{x}\frac {\partial} {\partial y},
\end{equation}
where $m^{*}$ is the electron's effective mass, $V(y)$ the stationary potential, $\sigma_{x}$ one of Pauli's matrices and $\alpha_{R}$ is Rashba's parameter which is space-dependent in this model and brings the problem of having a non hermitic hamiltonian. This is easily solved by realizing the following symmetrization
\begin{equation}
\hat{H}_{so}=-i\sigma_{x}\alpha_{R}(y)\frac {\partial} {\partial y}+\frac {1}{2}\frac {\partial} {\partial y}\alpha_{R}(y).
\end{equation}
Therefore the Schr\"{o}dinger time-dependent equation with Rashba's spin orbit coupling for this model is:
\begin{eqnarray}
[(-\frac {\hbar^{2}} {2m^{*}} \frac {d^{2}} {dy^{2}}+V(y))\mathbb{I}-i\sigma_{x}(\alpha_{R}(y)\frac {\partial} {\partial y}+\nonumber\\
\frac {1} {2}\frac {\partial}{\partial y}\alpha_{R}(y))]|\Psi(y,t)\rangle=i\hbar|\Psi(y,t)\rangle,
\end{eqnarray}
with $\mathbb{I}$ being the identity matrix and $|\Psi(y,t)\rangle$ a spinor of the form:
\begin{equation}
|\Psi(y,t)\rangle=\left(
\begin{array}{c} \Psi_{\uparrow}(y,t)\\\Psi_{\downarrow}(y,t)\end{array}\right).
\end{equation}
Also by using the time evolution equation, which for a conservative system is:
\begin{equation}
|\Psi(y,t)\rangle=e^{-i\frac {(t-t_{0})\hat{H}} {\hbar} }|\Psi(y_{0},t_{0})\rangle,
\end{equation}
with the hermitic operator $e^{-i\frac {(t-t_{0})\hat{H}} {\hbar} }$, and $\Psi(y_{0},t_{0})$ being the wave function with centroid $y_{0}$ in time $t_{0}$ and $t_{0}<t$.
Initial and boundary conditions
\vspace{3mm}\\
In order to realize a numerical simulation of Schr\"{o}dinger's equation incorporating R-SOI a gaussian spin polarized wavepacket of the next form is considered:
\begin{equation}
|\Psi(y,0)\rangle=\Psi(y,0)\xi_{\uparrow,\downarrow},
\end{equation}
where $\xi_{\uparrow,\downarrow}$ gives the packet's polarization and
\begin{equation}
\Psi(y,0)=\sqrt{\frac {2\pi} {\Delta y}}e^{ik(y-y_{0})}e^{-\frac {(y-y_{0})^{2}} {2\Delta y^{2}}},
\end{equation}
is a wavepacket that moves to the right with an initial momentum of  $p=\hbar k$ and a packet spread of $\Delta y$.

The boundary conditions are set considering  a packet confined inside a one-dimensinal box of length $L$, so that the wave functions for each spin state equal zero at all time for
\begin{equation}
\left(
\begin{array}{c} \Psi_{\uparrow}(L,t)\\\Psi_{\downarrow}(L,t)\end{array}\right)=\left(
\begin{array}{c} \Psi_{\uparrow}(0,t)\\\Psi_{\downarrow}(0,t)\end{array}\right)=\left(
\begin{array}{c} 0\\0\end{array}\right).
\end{equation}

This conditions are satisfied for certain values of $\Delta y$ and the initial position of the packet's centroid. In order to consider negligible the wave functions in the boundary is necessary realize an adequate selection of the simulation time, the incident energy and the packet's spread. Those conditions have been previously summarized by Golberg \cite{Goldberg}.

\subsection{Instantaneous measurement of the energy}

For us to be able to realize energy measurements during the temporal evolution of the wavepacket is necessary to use the definition of the expectation value of the hamiltonian:

\begin{equation}
\langle\hat{H}\rangle= \int_{-\infty}^{ \infty} \Psi^{*} (y) \hat{H} \Psi (y) dy
\end{equation}

By using the finite difference method \cite{Marisol} this definite integral can be expressed as a Riemann sum obtaining for each state of the wavepacket:

\begin{eqnarray}
\langle\hat{H}_{\uparrow,\downarrow}^{n}\rangle=\delta y\Sigma_{j=0}^{J}[ \Psi_{\uparrow,\downarrow j}^{*n}(-\frac {\hbar^{2}} {2m^{*}}\frac {1} {\delta y^{2}}(\Psi_{\uparrow,\downarrow(j+1)}^{n}+\Psi_{\uparrow,\downarrow(j-1)}^{n}\nonumber\\
-2 \Psi_{\uparrow,\downarrow j}^{n})+V_{j}\Psi_{\uparrow,\downarrow j}^{n}-\frac {i} {\delta y}\alpha_{j}(\Psi_{\downarrow,\uparrow(j+1)}^{n}-\Psi_{\downarrow,\uparrow j}^{n})\nonumber\\
-\frac {i} {2}(\frac {\alpha_{j+1}-\alpha{j}} {\delta y})\Psi_{\downarrow,\uparrow j} )]\nonumber\\
\end{eqnarray}

 The above expression indicates the expectation value of the hamiltonian for each instant of time $n$. So we will define the energy deviation of the wave in every instant of time $n$ as:

\begin{equation}
 \label{DE}
\Delta E_{\uparrow,\downarrow}=|E_{inc}-\langle\hat{H}_{\uparrow,\downarrow}^{n}\rangle|
\end{equation}

\section{Applicability of the TEUR}

In this section the energy deviation of a spin polarized electronic wavepacket through a quasi one dimensional $GaAs/AlAs/GaAs$ heterostructure which incorporates a stationary potential barrier with R-SOI is shown and discussed. The simulation is realized for two types of barrier. The thin barrier being of $3${\AA} width and the thick barrier being of $5${\AA} width.
The input parameters have been summarized in table 1, leaving only to variation the incident wavepacket's energy $E_{inc}$ and the values of Rashba's parameter $\alpha$. Both values will be specified on each figure.

\begin{table}
\label{tab:table1}
\caption{This table summarizes the values of the different parameters used in the simulation, and the units used for each of them}
\begin{ruledtabular}
\begin{tabular}{cccl}
Symbol&Value&Units&Description\\
\hline
$\xi_{\uparrow}$&$\frac {1} {\sqrt{2}}\left(\begin{array}{c} 1\\i\end{array}\right)$&-----& wavepacket polarization\\
$\delta y$& $.4$ & {\AA} & Spatial partition \\
$\delta t$& $1$ & fs & Temporal partition \\
$V$& $.4$ & $eV$ & Barrier's height \\
$L$& $1200$ & {\AA} & System Length \\
$J$& $3000$ &-----& Number of spatial partitions \\
$N$& $50$ &-----& Number of temporal partitions \\
$\Delta y$& $100$ & {\AA} & Packet spreading \\
$m_{e}$& $.05677$ & $\frac {eV fs^{2}} {{\AA}^{2}}$ & Electron's mass \\
$m^{*}_{GaAs}$\footnotemark[1]& $0.067m_{e}$ & $\frac {eV fs^{2}} {{\AA}^{2}}$ & GaAs effective mass \\
$m^{*}_{AlAs}$\footnotemark[1]& $0.15m_{e}$ & $\frac {eV fs^{2}} {{\AA}^{2}}$ & AlAs effective mass \\
$\hbar$& $0.6582$ & $eV fs$ & Planck's constant \\
$c$& $3\times10^{4}$ & $\frac {{\AA}} {fs}$ & Speed of light \\
\end{tabular}
\end{ruledtabular}
\footnotetext[1]{Values obtained from , \cite{Vurga}.}
\end{table}

\subsection{Thin barrier 3{\AA}}

Firstly the energy deviation of a wavepacket during its transmission trough the heteroestructure without taking into account R-SOI is analyzed. The parameters not shown in the text will be described in the proper figure.

\begin{figure}
 \hspace*{-8mm}\includegraphics[width=3.4in,height=3.4in]{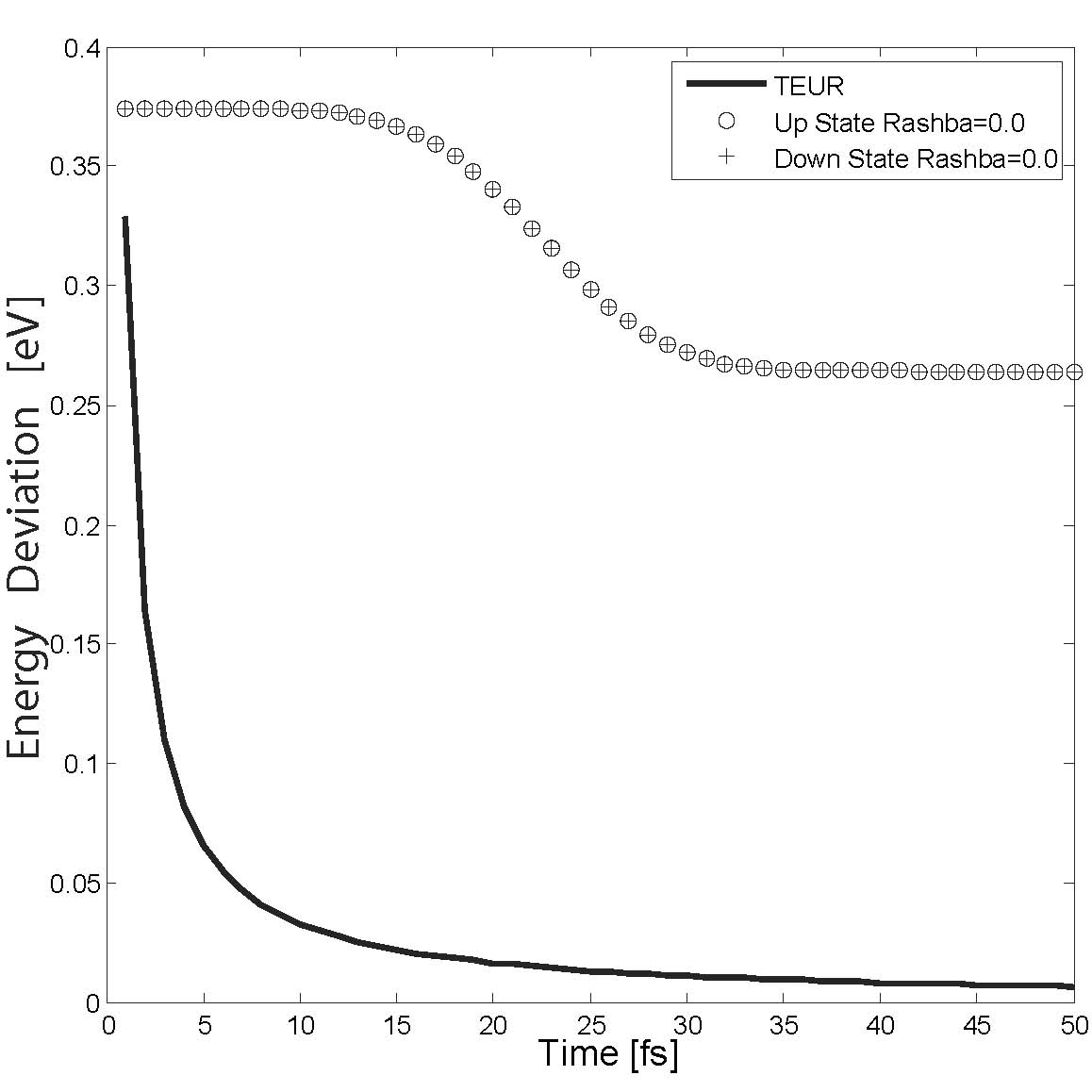}\\
 \caption{\label{Fig1}}
\end{figure}

Figure \ref{Fig1} shows how in the case where $E_{inc}>V$ without R-SOI both spin states overlap determining a state degeneracy by the spin during the whole simulation. The TEUR is also shown in Figure \ref{Fig1} and it will be shown in all of the next figures and it's the curve that shows the theoretical limit where equation \ref{DE} is evaluated. As it was expected the plot with no R-SOI lands on the conservation zone . This behavior will be observed as tendency in most of the figures.

\begin{figure}
 \hspace*{-8mm}\includegraphics[width=3.4in,height=3.4in]{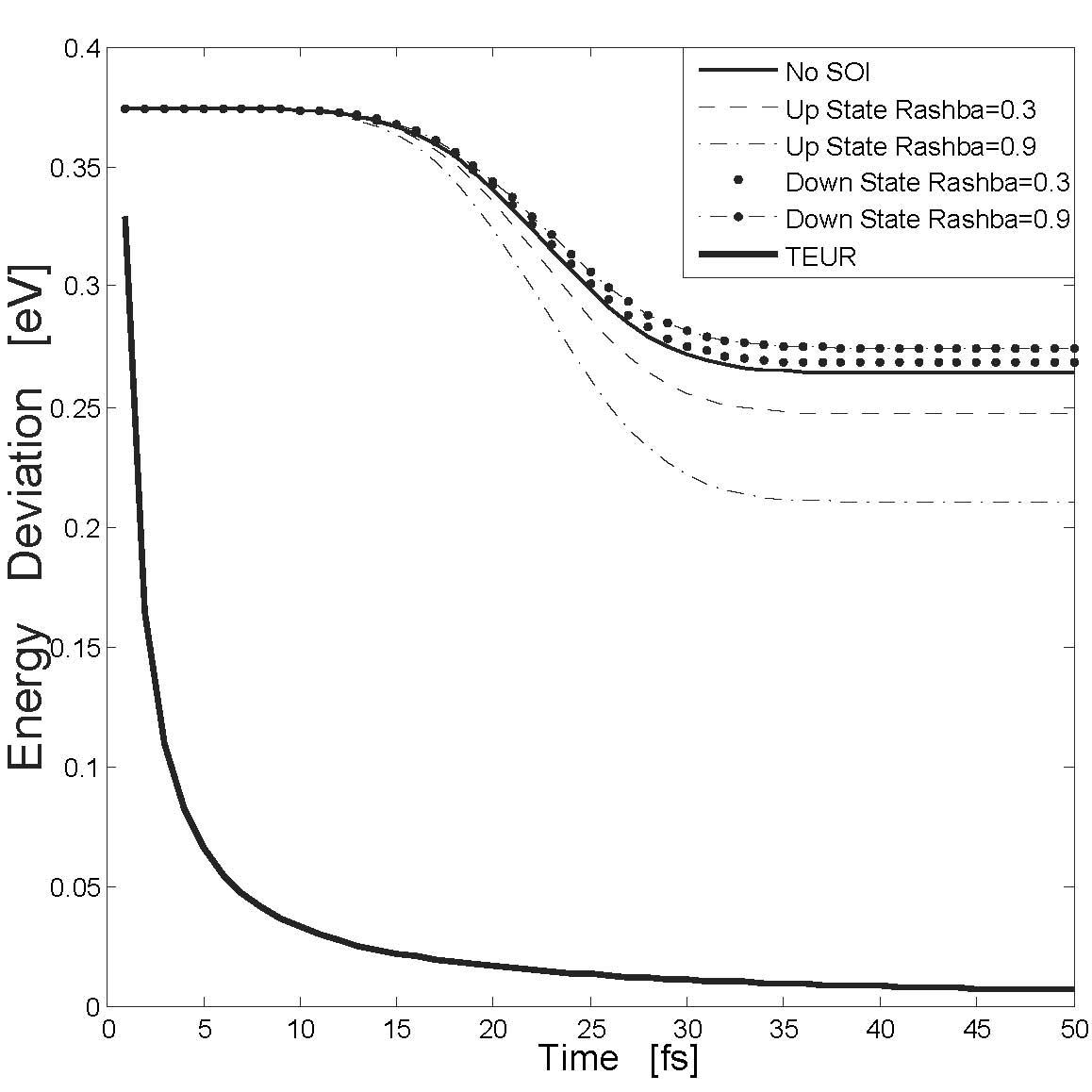}\\
 \caption{\label{Fig2}}
\end{figure}

Figure \ref{Fig2} shows us the same case discussed in Figure \ref{Fig1}, however; in this case Rashba's parameter varies, showing how its incorporation into the system causes an unfolding of the spin states, indicating no degenerated states when the packet comes out of the barrier, what exhibits a difference in the wave function associated to the different states of the electron. The continuous line shows the case without R-SOI, the dashed lines show the up state for different values of Rashba's parameter while the other two point-like lines correspond to the down state. It's noticed how Rashba's parameter does not cause a symmetric split of the spin states, in the sense that the energy deviation of the up and down states are not evenly separated from the case without R-SOI, this because of the packet's initial polarization. The degeneracy break is an important result because indicates how R-SOI allows us to select the spin states acting as a filter, though, R-SOI allows us to manipulate the spin states according to the packet's initial polarization. This result gives sense to speaking about R-SOI and the uncoupling caused by it, as a spin filter for both cases of the quantum dispersion of the electronic packet.

\begin{figure}
 \hspace*{-8mm}\includegraphics[width=3.4in,height=3.4in]{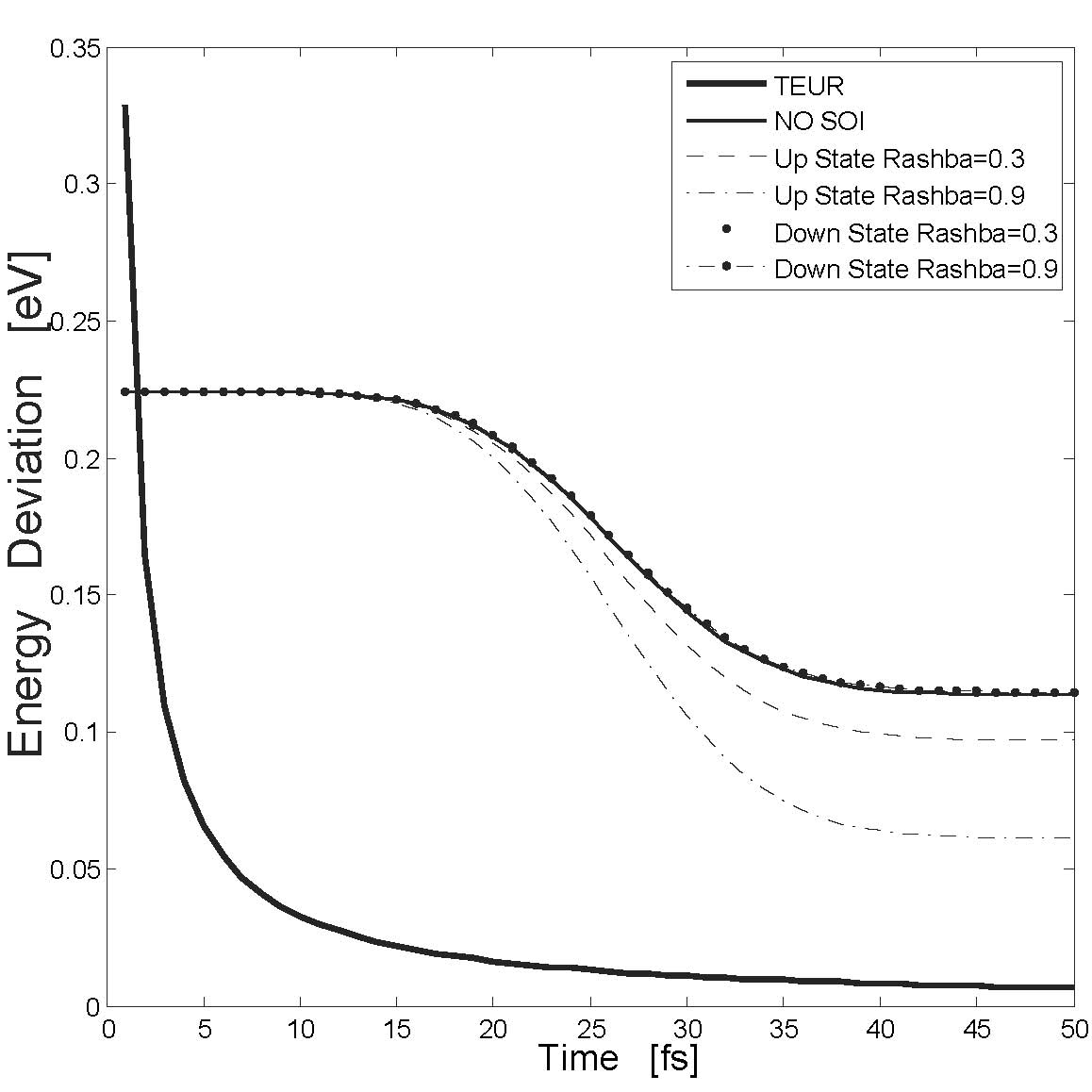}\\
 \caption{\label{Fig3}}
\end{figure}

It is evident from Figure \ref{Fig3} that the energy deviations diminish for the tunneled case where $E_{inc}<V$, however the same behavior that uncouples the spin states is observed for different values of Rashba's parameter. By noticing that for a barrier of 3{\AA} the R-SOI allows us to manipulate the spin states according to the packet's initial polarization, showing  the behavior a spin sensor should have for both cases of the energy deviation $E_{inc}>V$ and $E_{inc}<V$.

\subsection{Thick barrier $5${\AA}}

The case here presented corresponds to a barrier with a $5${\AA} width. In the same way as before the parameters not given in the text will be described explicitly on each figure.

\begin{figure}
 \hspace*{-8mm}\includegraphics[width=3.4in,height=3.4in]{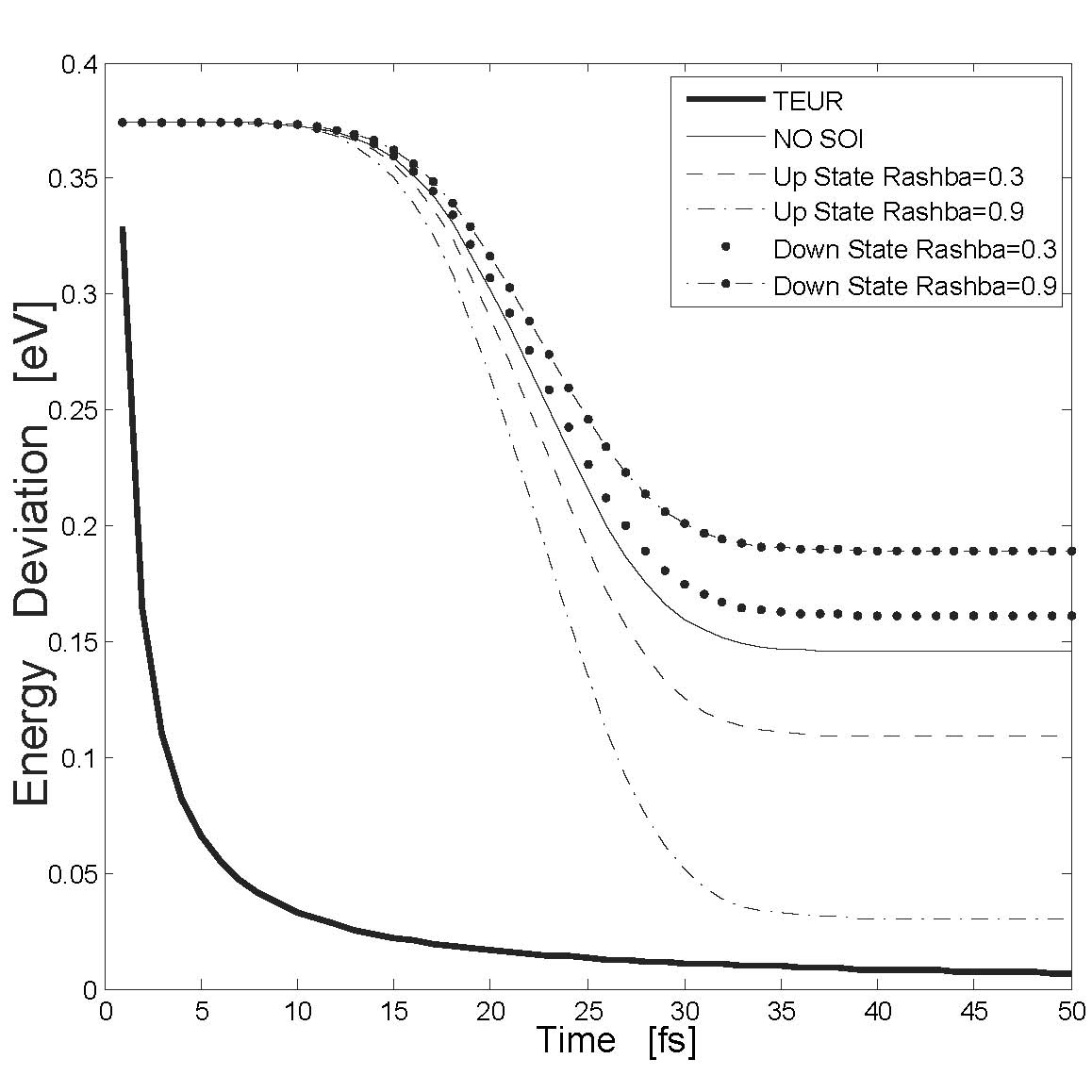}\\
 \caption{\label{Fig4}}
\end{figure}

The transmitted case, i.e. $E_{inc}>V$, without R-SOI will be analyzed first. Figure \ref{Fig4} shows how the spin states overlap again in this case as they did in the thin barrier case.

\begin{figure}
 \hspace*{-8mm}\includegraphics[width=3.4in,height=3.4in]{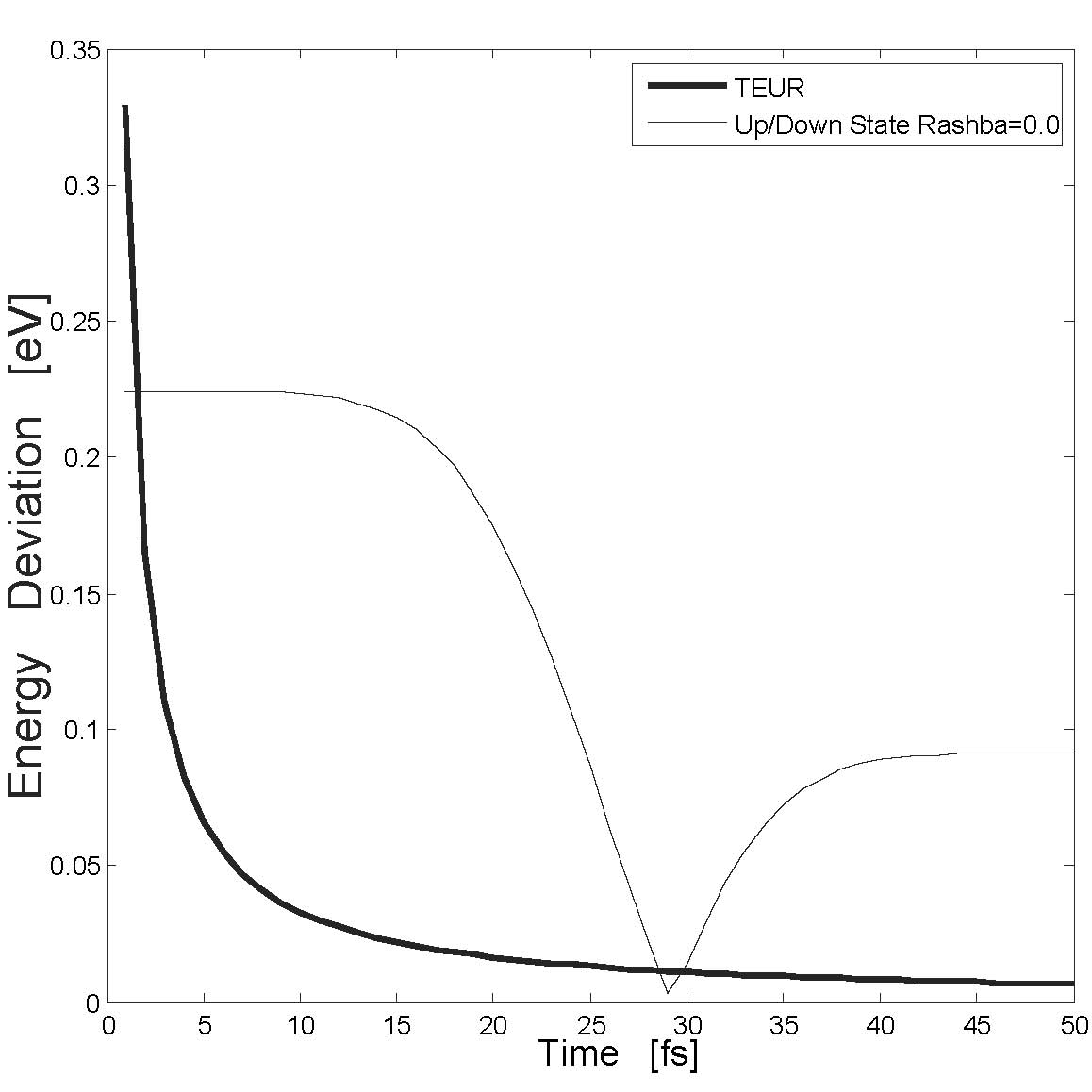}\\
 \caption{\label{Fig5}}
\end{figure}

Figure \ref{Fig5} shows us how when incorporating R-SOI a break on degeneracy of the spin states appears and this states are uncoupled having the down states landing above the curve where no R-SOI is considered. However, it may be appreciate that for the case where the Rashba parameter equals $0.3 eV{\AA}$  the gap between this curve and the one with no R-SOI is bigger compared to the same case with a thin barrier. This behavior allows us to infer, an increment in the intensity of the R-SOI, which can be deduced from the term $H_so(y)$ of equation (\ref{Rashba}) because the packet had to interact with a thicker barrier.

\begin{figure}
 \hspace*{-8mm}\includegraphics[width=3.4in,height=3.4in]{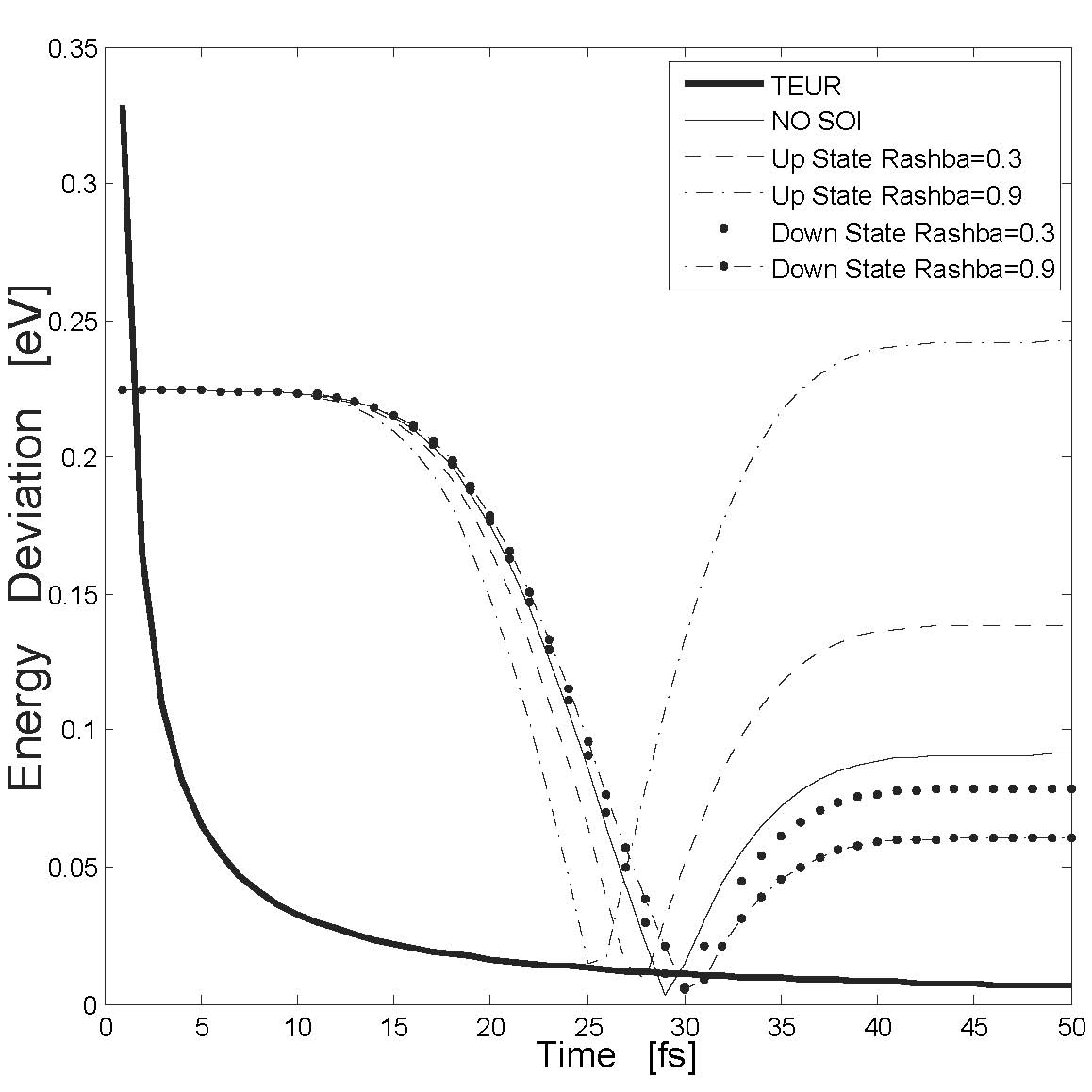}\\
 \caption{\label{Fig6}}
\end{figure}

The other case studied is the case when $E_{inc}<V$, i.e. the tunneled case. Following the same scheme  figure \ref{Fig6} shows how a spin state degeneracy exists in the case without R-SOI. However, the barrier's width provokes  an abrupt change in the behavior of the energy deviation during the wavepacket's evolution. This change will be discussed  when finalizing this section.
When taking into account Rashba's parameter into the simulation, once again it is observed in figure \ref{Fig6} a rupture on degeneracy by the spin. However, this time the states are inverted, having the up states above the curve without R-SOI and the down states below it.
The reason why the behavior for a thick barrier is different can be explain by going back to our definition of energy deviation and remembering it includes the concept of absolute value. Which indicates for this case that during the time that the packet travels trough the barrier exists an instant where $\langle\hat{H}\rangle=E_{inc}$ causing the abrupt change observed in figure \ref{Fig5} and figure \ref{Fig6}. This indicates that under this range of paramteres, a $\Delta E=0$ would be calculated meaning an absolute accuracy. This opposes the limited determinism principle that has been assumed, explaining why the singular points land above of below the theoretical limit.

\begin{figure}
 \hspace*{-8mm}\includegraphics[width=3.4in,height=3.4in]{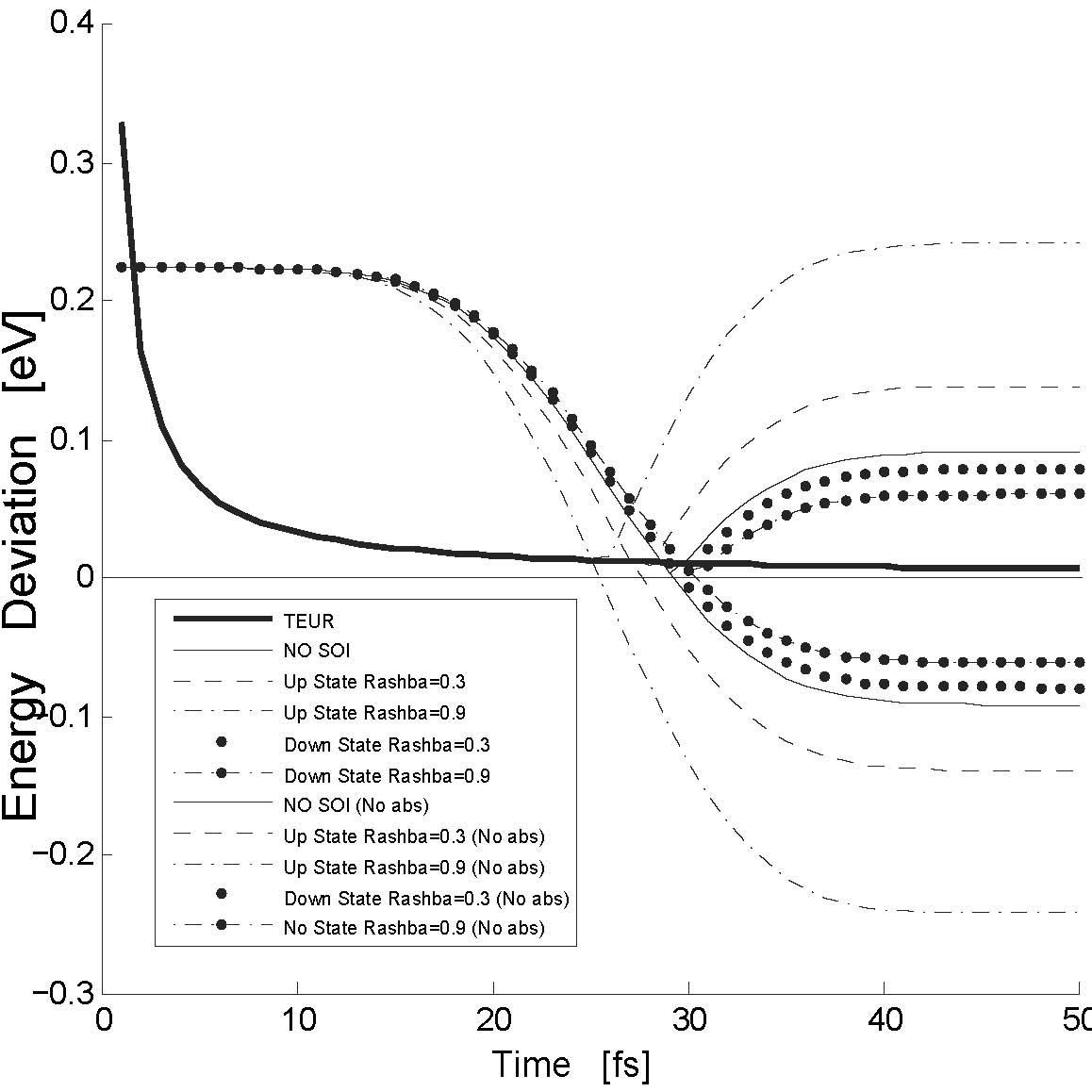}\\
 \caption{\label{Fig7}}
\end{figure}

Figure \ref{Fig7} shows the energy deviation without taking into account the absolute value. The part of the curves located in the negative value part of the plane correspond to deviations not associated with the limited determinism principle of equation (\ref{DE}), but they have been taken into account in order to explain the inversion of the spin states shown previously in figure \ref{Fig6}.

Clearly, the Fig.\ref{Fig7} shows how the smooth behavior still manifests without taking the absolute value of the energy deviation, and how departing from this graphs it could be possible to keep interpreting the R-SOI as the cause of the unfolding of the spin states and the possible applicability of this phenomena to spin filters.

\section{Concluding Remarks}
For the different polarized wavepackets under simulation it was shown that R-SOI creates a rupture of degeneracy of the spin states, as expected. This filtering effect can be neatly resolved by TEUR graphs, whose curves remains in the allowed zone (above the TEUR-limit border). Besides, it was observed that for the tunneled case and thick barrier ($5${\AA}) the plot does not show a smooth behavior like the one shown in the other cases, however this feature can be explained by the definition of the packet's energy deviation and realizing that this abrupt behavior was the cause of an inversion of the spin states and it seams to mean that there exists a short interval of time in which $\langle\hat{H}\rangle = E_{inc}$ indicating a violation to the TEUR-limit stated before.

\begin{acknowledgments} GBE is thankful for the financial support given by the Facultad de F\'{\i}sica, Universidad de la Habana, Cuba.
\end{acknowledgments}


\section*{References}
\begin{thebibliography}{14}
\bibitem{Olkhovsky} V. S. Olkhovsky, E.Recami,J. Jakiel, Physics Reports \textbf{398}, 133 (2004).
\bibitem{Giribet} G. E. Giribet, Rev. Mex. Fis. E \textit{51}, 23 (2006).
\bibitem{Aharonov} Y. Aharonov,D. Bohm, Phys. Rev. \textbf{122}, 1649 (1961).
\bibitem{Leo1} L. Diago-Cisneros, H. Rodr\'{\i}guez Coppola, R. Per\'{e}z-\'{A}lvarez, and P. Pereyra, Phys. Rev. B {\bf 74}, 045308 (2006).
\bibitem{Coppola}  H. Rodr\'{\i}guez-Coppola, L. Diago-Cisneros, R. Per\'{e}z-\'{A}lvarez, J. App. Phys {\bf 102}, 094315 (2007).
\bibitem{Datta} Datta and Das, App. Phys. Lett., {\bf 56}, 665 (1990).
\bibitem{Goldberg} A. Goldberg. H. M. Schey, Am. J. Phys {\bf 35}, 177.
\bibitem{Burgos} L. Burgos ``\textit{Estudio del acoplamiento sp\'{\i}n-\'{o}rbita en sistemas peri\'{o}dicos}'', Tesis de Licenciatura, UNAM-Ensenada, Baja California, M\'{e}xico (2005).
\bibitem{Marisol} M. Ochoa ``\textit{Estudio de la din\'{a}mica de los paquetes de ondas esp\'{\i}n polarizados en presencia de acoplamiento esp\'{\i}n-\'{o}rbita}'' Tesis de Licenciatura, UNAM-Ensenada, Baja California, M\'{e}xico (2006).
\bibitem{Bohm} D. Bohm, \textit{``Quantum theory''}, Dover Publications Inc. New Tork (1979).
\bibitem{Vurga} I. Vurgaftmana, J. R. Meyer, L. R. Ram-Mohan, J. App. Phys {\bf 89}, 5815 (2001).
\end{thebibliography}
\end{document}